\documentclass[twocolumn,aps,prb,amsmath,floatfix,showpacs]{revtex4}
\usepackage{graphicx}

\begin{document}
\title{Coulomb blockade and Kondo effect in the electronic structure of
Hubbard molecules connected to metallic leads: 
a finite-temperature exact-diagonalization study}
\author{H. Ishida$^{1,2}$ and A. Liebsch$^2$}
\affiliation{$^1$College of Humanities and Sciences, Nihon University, Tokyo, 
              156-8550, Japan\\
$^2$Peter Gr\"unberg Institute and Institute of Advanced Simulations,
Forschungszentrum J\"ulich, 52425 J\"ulich, Germany}
\date{\today}

\begin{abstract}
The electronic structure of small Hubbard molecules coupled between two
non-interacting semi-infinite leads is studied in the low bias-voltage limit. 
To calculate the finite-temperature Green's function of the system, each 
lead is simulated by a small cluster, so that the problem is reduced to that 
of a finite-size system comprising the molecule and clusters on both sides.
The Hamiltonian parameters of the lead clusters are chosen such that their embedding 
potentials coincide with those of the semi-infinite leads on Matsubara frequencies.
Exact diagonalization is used to evaluate the effect of Coulomb correlations 
on the electronic properties of the molecule at finite temperature. 
Depending on key Hamiltonian parameters, such as Coulomb repulsion, one-electron 
hopping within the molecule, and hybridization between molecule and leads, the 
molecular self-energy is shown to exhibit Fermi-liquid behavior or   
deviations associated with finite low-energy scattering rates. The method is shown 
to be sufficiently accurate to describe the formation of Kondo resonances inside 
the correlation-induced pseudogaps, except in the limit of extremely low  
temperatures. These results demonstrate how the system can be tuned between the 
Coulomb blockade and Kondo regimes.
\end{abstract}

\maketitle

\section{Introduction}
\label{sec_1}

Finite-size electron systems linked to non-interacting electron reservoirs have
been a topic of intense theoretical and experimental study because of their 
relevance to quantum dot systems and single-molecule devices. Depending on the 
importance of correlation effects induced by the electron-electron Coulomb repulsion, 
different types of theoretical approaches are employed. 
For weakly correlated systems, ballistic electron transport
is studied within the one-electron approximation such as density-functional theory
(DFT).\cite{Thygesen:03,Brandbyge:02,Ventra:01,Hirose:95,Wortmann:02}
On the other hand, strongly correlated systems are modeled by tight-binding
Hamiltonians with Hubbard- or Anderson-type interaction terms and various many-body
techniques are applied.
\cite{Meir:91,Hershfield,Meir:93,Werner:10,Smirnov:11,Goyer:11,Anda:08,
Asai:05,Oguri:01,Oguri:01b,
Oguri:05,Numata:09,Tanaka:10,Mitchell:11,Vernek:11,Misiony:12,Costi:10, 
Jacob:09,Jacob:10a,Jacob:10b,Karolak:11,Surer:12,Romero:11, 
Beenakker:91,Konig:03,Braig:05,Muralidharan:06,Song:07} 
Two noticeable effects beyond the one-electron approximation are the Kondo effect
and Coulomb blockade, both of which are observed in quantum dot systems. 
\cite{Goldhaber:1998,Cronenwett,Schmid}
More recently, the Kondo effect was also observed in adsorbed molecules by scanning
tunneling spectroscopy and high-resolution photoemission
spectroscopy.\cite{Tsukahara,Minamitani,Mugarza:12,Ziroff:12}  
The Kondo effect in nano-size systems was studied theoretically by using the numerical 
renormalization group (NRG) technique for a variety of cases, such as multi-dot or 
multi-level systems and dots coupled to superconducting leads,\cite{Oguri:05,
Numata:09,Tanaka:10,Mitchell:11, Vernek:11,Misiony:12} within DFT combined with the
one-crossing approximation\cite{Jacob:09,Jacob:10a,Jacob:10b,Karolak:11} and the
continuous-time quantum Monte Carlo technique.\cite{Surer:12}
Coulomb blockade effects seen in electron transport through a finite-size 
interacting system are investigated by using rate-equation techniques and 
non-equilibrium Green's function theory.
\cite{Beenakker:91,Konig:03,Braig:05,Muralidharan:06,Song:07}

The aim of the present work is to introduce a new scheme for the investigation 
of quantum dots that is applicable in the full range between Kondo physics and 
Coulomb blockade, except in the limit of extremely low Kondo temperatures. To
illustrate this approach, we focus on small interacting molecules coupled to 
non-interacting semi-infinite electron reservoirs. The many-body properties of 
these systems are evaluated by using exact diagonalization (ED) at finite 
temperatures.\cite{Caffarel:94,Perroni:07,Liebsch:12} In order to apply ED, 
the semi-infinite leads are simulated by finite-size clusters. For a given chemical 
potential, the tight-binding Hamiltonian parameters of these clusters are 
chosen such that the difference between the surface-site Green's function of a 
semi-infinite lead and the corresponding cluster lead is minimized along 
the Matsubara axis. The finite-temperature Green's function of the total system 
consisting of molecule and lead clusters is then evaluated exactly within ED. 
Since the effective lead--cluster Hamiltonian is extremely sparse, at typical 
temperatures of interest only a limited number of excited states needs to be 
evaluated. Here we consider Hubbard 
chain and ring molecules attached to two metallic leads. In these systems 
lead clusters consisting of only five bath levels can accurately mimic 
the true embedding potentials down to $T\agt t_M/500$, where $t_M$ denotes 
the hopping parameter representing the metallic leads. Thus, the formation of 
Kondo resonances within the correlation-induced pseudogaps can be investigated. 
In the case of single leads, the problem reduces to the single-impurity Anderson
model. The cluster size can then be significantly increased so that much lower 
temperatures can be reached. An additional advantage of our method is that the 
zero-bias-voltage limit of the interacting system can be studied for arbitrary 
values of the Coulomb repulsion and molecule-lead hybridization, without further 
approximations.

The discretization of the semi-infinite leads and the application of ED to the
resulting finite-size system is analogous to the use of ED as impurity solver 
within the context of dynamical mean-field theory (DMFT) and its cluster 
extensions\cite{Georges:96,Kotliar:01,recent.dmft} where the Weiss mean-field 
is represented by a finite number of non-interacting levels. Multi-orbital as 
well as multi-site correlations have been studied for a variety of materials,
\cite{Liebsch:12} including various surfaces and heterostructures.
\cite{Ishida} The main difference is that in the present case the leads are 
assumed to be non-interacting, so that the self-consistent iterative procedure
is absent.   

An important feature of the ED approach is that it provides complete dynamical 
information, in particular, transfer of spectral weight between low and high 
excitation energies, formation of Hubbard bands, and opening of correlation-induced 
pseudogaps. A quantity of central interest therefore is the molecular self-energy 
which exhibits strong variations as a function of Hamiltonian parameters and 
temperature. In particular, Fermi-liquid behavior in the Kondo regime and 
correlation-induced finite scattering rates in the Coulomb blockade regime can 
clearly be identified.
  
The outline of this paper is as follows. In Section \ref{sec_2} we describe our
theoretical model for the molecule--lead system and discuss several details of 
the calculation of the molecular self-energy and interacting Green's function. 
In Section \ref{sec_3} we present the numerical results and the discussion, 
with special emphasis on the Coulomb blockade and Kondo effect. 
Section \ref{sec_4} contains the summary. In the Appendix we consider a single 
adatom on a semi-infinite lead and examine the temperature range in which a 
finite-size cluster can
be used to simulate a semi-infinite lead. Since this case is equivalent to the 
single-impurity Anderson model, the results can directly be compared with those 
of other schemes that are applicable at arbitrarily low temperatures.

\section{Theory}
\label{sec_2}

\subsection{Formalism}
\label{sec_2A}

\begin{figure}[t] 
\begin{center}
\includegraphics[width=0.33\textwidth]{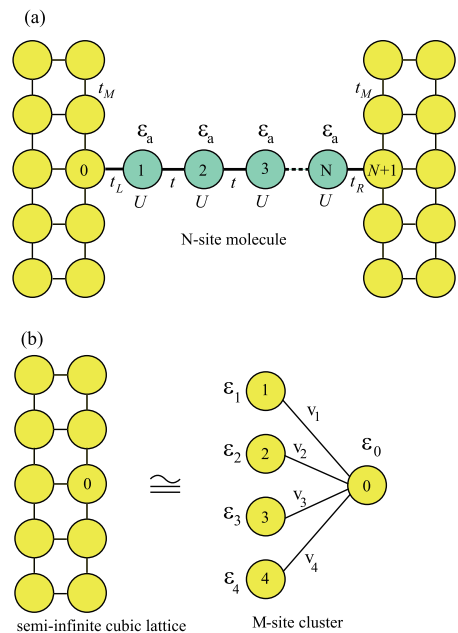}
\end{center}
\caption{\label{Fig_md}
(a) Tight-binding model for an $N$-site molecule attached to two non-interacting
semi-infinite leads. (b) For the evaluation of the Green's function of the molecule, 
the two semi-infinite leads are replaced by a cluster consisting of $M$ levels,
as shown here schematically for the left lead.}
\end{figure}

We consider a molecule consisting of $N$ atomic sites and linked to two 
semi-infinite metal leads, as shown schematically in Fig.\ \ref{Fig_md}(a).
The isolated molecule is modeled by a single-site
Hubbard Hamiltonian characterized by the onsite energy $\epsilon_a$, the
nearest-neighbor hopping interaction $t$, and the onsite Coulomb repulsion $U$,
\begin{equation}
\hat{h}_C=\sum_{i \sigma} \epsilon_a\ \hat{n}_{i \sigma}
- \sum_{\langle i,j\rangle \sigma} t\ \hat{c}_{i\sigma}^\dagger\hat{c}_{j\sigma},
+ \sum_{i} U\ \hat{n}_{i\uparrow}\hat{n}_{i\downarrow},
\label{eq1}
\end{equation}
where $\hat{c}^\dagger_{i\sigma}$ ($\hat{c}_{i\sigma}$) creates (destroys) an
electron with spin $\sigma$ at site $i$  ($1\leq i \leq N$),
$\hat{n}_{i\sigma}=\hat{c}^\dagger_{i\sigma} \hat{c}_{i\sigma}$, and the
summation in the second term is taken over pairs of nearest-neighbors.
Hereafter, we adopt the notation where the matrix (operator) corresponding to a
quantity $A$ is denoted by $\hat{A}$ while its matrix elements are defined as $A_{ij}$.
For simplicity, we limit the discussion here to one level per molecular site and 
purely onsite Coulomb interactions. Equivalently, it would also be feasible to 
investigate multi-orbital interactions, including inter-orbital Coulomb and Hund's
rule coupling, for instance, in transition metal ions attached to semi-infinite 
leads. 
Throughout this 
paper, the hopping integral within the molecule is taken as unit of the energy scale, 
i.e., $t=1$.

The left (right) lead is represented by non-interacting electrons on a semi-infinite
simple cubic lattice with nearest-neighbor hopping interaction $t_M$ and the onsite
energy level is chosen as zero of the energy scale:
\begin{equation}
\hat{h}_{L(R)}=
- \sum_{\langle i,j\rangle \sigma} t_M\ \hat{c}_{i\sigma}^\dagger\hat{c}_{j\sigma},
\label{eq2}
\end{equation}
where $i,j<1$ ($i,j>N$), so that the energy bands of both leads exhibit a finite density
of states (DOS) in the energy range $[-6t_M, 6t_M]$.

The molecule is linked to the left (right) lead via the hopping integral between site 1
($N$) of the molecule and site 0 ($N+1$) of the left (right) lead.
The mixing term of the Hamiltonian describing the molecule--lead hybridization is 
expressed as
\begin{equation}
\hat{h}_{mix}=- \sum_\sigma  \left(t_L \hat{c}^\dagger_{0\sigma}\hat{c}_{1\sigma}+
t_R \hat{c}^\dagger_{N+1\sigma}\hat{c}_{N\sigma}\right)+{\rm h.c.},
\label{eq3}
\end{equation}
where, for simplicity, the hopping integrals on both sides are assumed to coincide:
$t_L = t_R$. The Hamiltonian of the total system consisting of the molecule and the
two leads is given by
\begin{equation}
\hat{H}=\hat{h}_C+\hat{h}_L+\hat{h}_R+\hat{h}_{mix}. \label{eq4}
\end{equation}

We investigate the electronic structure of the molecule described by Eq.\ (\ref{eq4}) 
for a wide range of Hamiltonian parameters. As we consider the low bias-voltage limit, 
both leads have the same chemical potential which is denoted as $\mu$. 
The Green's function $\hat{G}$ of the molecule can be written as
\begin{eqnarray}
G_{ij}(i\omega_n)&=&\Bigl[i\omega_n+\mu-\hat{h}_C^0
-\hat{\Sigma}(i\omega_n)\nonumber \\
&-&\hat{s}^L(i\omega_n+\mu)-\hat{s}^R(i\omega_n+\mu)
\Bigr]_{ij}^{-1},\label{eq5}
\end{eqnarray}
where $1\leq i, j\leq N$, $\omega_n=(2n+1)\pi T$ ($n\geq 0$) are Matsubara
frequencies at temperature $T$, $\hat{h}_C^0$ denotes the first two terms of
Eq.\ (\ref{eq1}), and $\hat{\Sigma}(i\omega_n)$ is the self-energy matrix accounting
for electron correlation effects within the molecule. As we also allow for interatomic
Coulomb correlations, the self-energy matrix has off-diagonal components with respect
to site index. This approach differs from recent ones\cite{Jacob:09,Jacob:10a,
Jacob:10b,Karolak:11} in which the self-energy of each site is assumed to be local
and determined in a self-consistent manner similarly to the layer DMFT 
approach.\cite{Potthoff:99} In the present work, we consider only
paramagnetic solutions and omit the spin index $\sigma$ hereafter.
In Eq.\ (\ref{eq5}),  $\hat{s}^{L(R)}$ denotes the embedding potential describing
the one-electron hybridization effects due to the left (right) lead on the 
molecule.\cite{Inglesfield:01,Ishida:09} These embedding potentials give rise to
broadening and shifting of the molecular levels and therefore play the role of 
contact self-energies.\cite{Asai:05}

For the present geometry, only the $\lbrace 11\rbrace$ element of $\hat{s}^L$ 
is non-vanishing:
\begin{equation}
s^L_{11}(z)=t_L^2\left[z-\hat{h}_L\right]^{-1}_{00}=t_L^2\, g_{00}(z). \label{eq6}
\end{equation}
Similarly, the only non-vanishing element of $\hat{s}^R$ is
\begin{equation}
s^R_{NN}(z)=t_R^2\left[z-\hat{h}_R\right]^{-1}_{N+1,N+1}
 =t_R^2 \, g_{N+1,N+1}(z).  \label{eq7}
\end{equation}
The surface Green's functions appearing in these expressions are given by 
\begin{equation}
g_{ii}(z)=\int_{-6t_M}^{6t_M}\ d\epsilon\
\frac{\rho_i(\epsilon)}{z-\epsilon},\label{eq8}
\end{equation}
where $\rho_i(\epsilon)$ denotes the local density of states per spin at surface 
site $i$ within the left ($i=0$) or right ($i=N+1$) lead.

In order to make use of ED for the evaluation of the interacting Green's function 
of the molecule, Eq.\ (\ref{eq5}), 
we follow a procedure that has proved to be very useful in analogous
DMFT calculations. The surface Green's functions $g_{ii}(z)$ ($i=0,N+1$) 
representing the continuous spectra of the leads are approximated by those 
of finite clusters consisting of $M$ levels, as depicted in Fig.\ \ref{Fig_md}(b)
for the left lead. The $\{00\}$ element of the cluster Green's function is given by
\begin{equation}
    g_{00}^{cl}(z)=\left[z-\epsilon_0-\sum_{k=1}^{M-1}
 \frac{v_k^2}{z-\epsilon_k}\right]^{-1}, \label{eq9}
\end{equation}
where $\epsilon_k$ ($0\leq k\leq M-1$) and $v_k$ ($1\leq k\leq M-1$) are the 
energy levels and intra-cluster hybridizations, respectively. An analogous 
expression holds for the surface Green's function of the right lead, $g_{N+1,N+1}(z)$.
(Note that the cluster levels $\epsilon_k$ do not refer to actual lattice sites 
within the leads. Instead, they represent auxiliary quantities to simulate the 
spectral distributions of the leads.) 
The discretization of $g_{00}(z)$ is not suitable on the real energy axis since 
$g_{00}(z)$ has a continuous energy spectrum while $g^{cl}_{00}(z)$ possesses 
only a finite number of poles. Thus, Im\,$g_{00}^{cl}(z)\rightarrow 0$ or $-\infty$
in the limit $z\rightarrow 0$, whereas $g_{00}(z)$ for metallic leads remains finite, 
so that, in the very low-energy region, the cluster Green's function deviates 
strongly from the actual lead Green's function. These discrepancies are absent
if the calculation is restricted to finite temperatures. $g_{00}(z)$ can then 
accurately be fitted by $g^{cl}_{00}(z)$ at Matsubara frequencies, since both 
functions vary smoothly along the imaginary energy axis. As shown in 
Ref.~\onlinecite{Liebsch:12}, in finite-$T$ ED/DMFT calculations for typical 
multi-orbital materials, two or three bath levels per orbital are adequate 
to achieve adequate fits for temperatures in the range $T\approx W/50, \ldots, 
W/200$, where $W$ is the bandwidth. For the present case, this implies
$T\approx 0.025,\ldots,0.10$. To reach lower temperatures therefore requires 
accordingly larger lead clusters.            
 
As in standard ED/DMFT calculations, for a given chemical potential $\mu$, the 
discretization of $g_{00}(z)$ can be achieved by determining $\epsilon_k$ and 
$v_k$ in Eq.\ (\ref{eq9}) via minimization of the quantity\cite{Liebsch:12}
\begin{equation}
I=\sum_n  W_n \mid g_{00}(i\omega_n+\mu)-g^{cl}_{00}(i\omega_n+\mu) \mid^2,\label{eq10}
\end{equation}
where the weight function $W_n$ is chosen as $1/\omega_n$ in order to provide
greater accuracy at low $\omega_n$. (The large frequency behavior is less relevant
in this fit since both Green's functions approach $1/(i\omega_n)$ at large 
$\omega_n$.)  Other choices, such as $W_n=1$ or 
$W_n=1/\omega_n^2$, usually give very similar results, even though the auxiliary  
cluster parameters $\epsilon_k$ and $v_k$ may vary slightly.         
With decreasing temperature $T$ the lowest Matsubara frequency approaches the 
real energy axis, so that the fitting becomes less accurate. As will be
demonstrated in the next section and Appendix, the true lead Green's function can
be simulated by that of a relatively small 5-level cluster with sufficient accuracy
as long as the temperature is approximately in the range $T\agt t_M/800$.
Moreover, the lower boundary of this temperature range can be reduced by increasing
the cluster size. As a consequence, it is feasible to describe the Kondo effect
on the spectral density, if the associated Kondo temperature is comparable with
this temperature range.

It should be noted here that the Matsubara temperature used in the fitting of the 
lead surface Green's function may be viewed as a fictitious temperature $T_M$ that does 
not need to coincide with the physical temperature $T$. Instead, its choice is mainly 
determined by the number of cluster levels used to simulate the semi-infinite leads. 
Evidently, a larger value of $M$ permits fitting at lower values of $T_M$.     
This point will be addressed further in the Appendix where an extremely small 
value of $T_M$ is chosen for the evaluation of the self-energy of a single adatom 
over a wide range of real temperatures. In most applications discussed below the 
Matsubara temperature is taken to be the physical temperature. 
 
Let us denote the non-interacting Green's function of the molecule linked to the two
semi-infinite leads by $\hat{G}^0$ and that linked to the two clusters by
$\hat{G}^{0,cl}$, where the term `non-interacting' signifies $U=0$ in the molecule. 
The interacting counterparts are $\hat{G}$ and $\hat{G}^{cl}$, respectively.
When the tight-binding parameters of the lead clusters are optimized as described 
above, one can presume that $\hat{G}^0(i\omega_n)\approx\hat{G}^{0,cl}(i\omega_n)$ at 
all Matsubara frequencies. As a result, when the Coulomb interaction in the molecule is
switched on, $\hat{G}(i\omega_n)$ should nearly coincide with $\hat{G}^{cl}(i\omega_n)$. 
We may therefore employ $\hat{G}^{cl}$ as a reasonable representation of the true
interacting Green's function, $\hat{G}$, of the molecule attached to the two 
semi-infinite leads. Schematically, the procedure outlined above proceeds via 
the following steps:
\begin{equation}
\hat{g} \approx \hat{g}^{cl} \rightarrow\hat{G}^{cl} \approx \hat{G} .  
\end{equation}
Below we do not distinguish between the molecular Green's functions $\hat{G}$ 
and $\hat{G}^{cl}$. We emphasize, however, that even if $\hat{G}$ agrees well 
with $\hat{G}^{cl}$ at Matsubara points, at real energies $\hat{G}^{cl}$ has 
a discrete level spectrum while that of $\hat{G}$ is continuous. 

The Hamiltonian of the interacting molecule linked to the two $M$-level clusters
is highly sparse. To compute $\hat{G}^{cl}(i\omega_n)$, we therefore make use 
of the Arnoldi algorithm which is ideally suited to evaluate the lowest eigen
states relevant at temperature $T$. $\hat{G}^{cl}(i\omega_n)$ is then derived 
via the Lanczos procedure for a finite number of excited states.   
In the present work, system sizes up to $n_s=N+2M=15$ have been investigated. 
More details concerning the  numerical procedure are provided in 
Refs.\ \onlinecite{Perroni:07} and \onlinecite{Liebsch:12}. (As mentioned earlier, 
the approach outlined above does not involve any self-consistency procedure 
as in DMFT since the leads are uncorrelated. Their fixed electronic
structure merely governs the boundary conditions of the interacting molecule.
The auxiliary lead quantities $\epsilon_k$ and $v_k$ characterizing the leads are 
therefore determined before carrying out the exact diagonalization.)  

\subsection{Semi-infinite vs. cluster leads}
\label{Sec_2B}

\begin{figure}[t] 
\begin{center}
\includegraphics[width=0.4\textwidth]{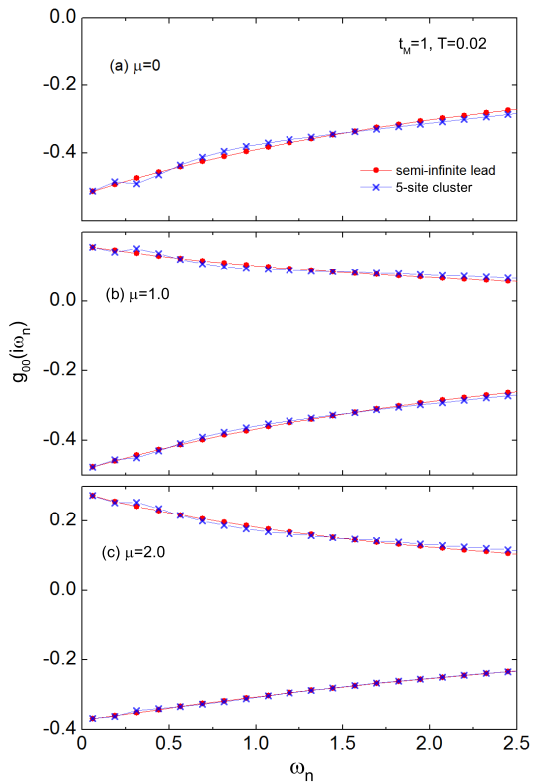}
\end{center}
\caption{\label{Fig_lg}
(Color online) Comparison of the $\{00\}$ element of the Green's function of a
semi-infinite lead (solid circles) and that of a 5-level cluster (crosses). 
(a) $\mu=0$, (b) $\mu=1$, and (c) $\mu=2$, for $t_M=1$ and $T=0.02$.
The real part of $g_{00}$ vanishes for $\mu=0$.}
\end{figure}

 As discussed above, the calculation of the 
electronic structure of an interacting molecule between semi-infinite leads is 
made feasible by simulating the $\{00\}$ ($\{N+1,N+1\}$) surface element of the 
Green's function of the left (right) lead in terms of a cluster Green's function, 
as indicated in Eq.\ (\ref{eq9}). To demonstrate the accuracy of this fitting
procedure, we compare in Fig.\ \ref{Fig_lg} both quantities as a function of 
$\omega_n$ for three values of $\mu$ for a semi-infinite lead with $t_M=1$
(band width $W=12$).

The lead clusters consist of five levels.
Thus, for $\mu\ne 0$, there are in total nine independent fit parameters:
$\epsilon_k$ ($0\leq k\leq 4)$ and $v_k$ ($1\leq k\leq 4$). At half-filling ($\mu=0$), 
this number is reduced to four because of symmetry reasons: $\epsilon_0=0$ and
the other four levels are symmetrically distributed with respect to $\epsilon=0$.
The fitting then becomes slightly less accurate than away from half-filling.
For the present choice of $T=0.02$, $g_{00}(i\omega_n)$ is seen to agree very well
with $g_{00}^{cl}(i\omega_n)$ in the whole $\omega_n$ range. At small Matsubara
frequencies weak cusps appear in the cluster Green's function as a result of its
singular behavior along the real energy axis.

\begin{figure}[t] 
\begin{center}
\includegraphics[width=0.4\textwidth]{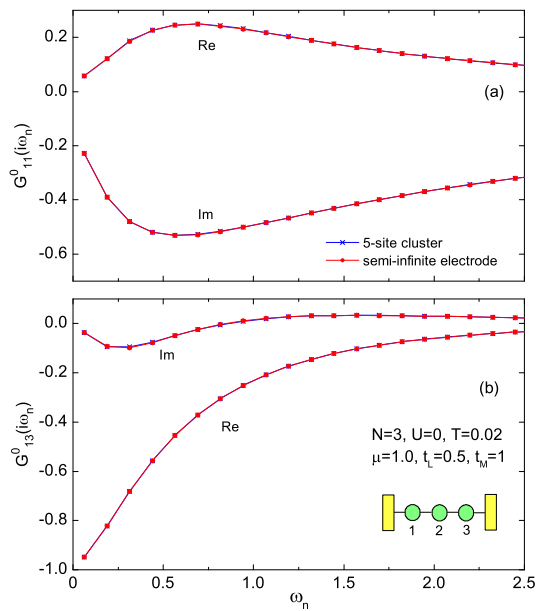}
\end{center}
\caption{\label{Fig_cG0}
(Color online) Non-interacting Green's function of a linear molecule with $N=3$
between two semi-infinite leads (solid circles) and that between two 5-level clusters
(crosses). (a) Diagonal $\{11\}$ element and (b) off-diagonal $\{13\}$ element,
for $t_M=1$, $t_L=0.5$, $\mu=1.0$,  and $T=0.02$.}
\end{figure}

Figure \ref{Fig_lg} suggests that the embedding potentials of the semi-infinite leads 
in Eq.\ (\ref{eq5}) can be approximated at Matsubara points by those of small clusters. 
To illustrate this point further, we compare in Fig.\ \ref{Fig_cG0} the
resultant non-interacting Green's function of a linear molecule ($N=3$)
between two semi-infinite leads, $G^0_{ij}(i\omega_n)$, with
the one of the same molecule between two 5-level clusters, 
$G^{0,cl}_{ij}(i\omega_n)$.  The Hamiltonian parameters correspond to
those in Fig.\ \ref{Fig_lg}(b) and the contact integrals are chosen as $t_L=0.5$. 
It is seen that both the diagonal and off-diagonal elements of $G^0_{ij}(i\omega_n)$ 
are in excellent agreement with the corresponding cluster elements 
$G^{0,cl}_{ij}(i\omega_n)$.

A crucial question determining the usefulness of the approach outlined above
concerns the range of temperatures in which accurate results can be obtained 
for a given cluster size. To explore this point, we present in the Appendix a 
careful study of the electronic structure of a single correlated adatom on a 
semi-infinite lead. In this case, a large range of cluster sizes can be employed
in order to systematically investigate the behavior of the self-energy at very 
low temperatures. Cluster sizes up to $M=11$ were used for $T\agt t_M/1600$.  
The results for 5-site clusters are found to agree quantitatively with those 
of larger clusters for $T\agt t_M/500$, and qualitatively for $T\agt t_M/800$.
These results, together with the ones shown in Figs.~\ref{Fig_lg} and \ref{Fig_cG0}, 
demonstrate the usefulness of our strategy of evaluating the Green's function of the
Hubbard molecule by simulating the semi-infinite metallic leads in terms of finite
clusters. In the following, we present results for the electronic structure of various
linear and ring molecules, where the true leads are replaced by clusters consisting
of five levels. 

\subsection{Spectral information} 

To demonstrate how the molecular electronic structure undergoes a transition
between the Kondo and Coulomb blockade regimes, we consider in the next section 
the partially integrated quasiparticle density of states which is defined as   
\begin{equation}
\bar\rho_i(\mu) = -  G_{ii}(\tau=\beta/2) ,
\end{equation}
where $G_{ii}(\tau)$ is the diagonal component of the imaginary-time Green's 
function at site $i$ ($\beta=1/T$):
\begin{eqnarray}
   G_{ii}(\tau) &=& \int d\omega\tilde{\rho}_i(\omega) \frac{e^{-\omega\tau}}
                    {1 +  e^{-\omega\beta}} \nonumber \\ 
    &=& -\sum_n e^{-i\omega_n \tau}\ G_{ii}(i\omega_n),
\end{eqnarray}
and the interacting quasiparticle DOS is defined by
\begin{equation}
\tilde{\rho}_i(\omega)=-\frac{1}{\pi}{\rm Im}G_{ii}(\omega+i\delta),
\label{eq14}
\end{equation}
with a positive infinitesimal $\delta$.
$\bar\rho_i(\mu)$ may therefore be expressed as
\begin{eqnarray}
\bar\rho_i(\mu) &=& -\sum_n e^{-i\omega_n \beta/2}\ G_{ii}(i\omega_n)\nonumber \\
   &=&  \int_{-\infty}^{\infty} d\omega\ F(\omega)\ \tilde{\rho}_i(\omega),
\label{eq15n}
\end{eqnarray}
where $F(\omega)=0.5/\cosh(\beta \omega/2) = (-T{\partial f}/{\partial\omega})^{1/2}$ 
is a distribution of halfwidth $w = 4\ln (2+\sqrt{3})T = 5.268 T$ centered about 
$\omega=0$ ($f$ is the Fermi function). Thus, at a given value of the chemical
potential $\mu$, $\bar\rho_i(\mu)$ represents the quasiparticle DOS of site $i$ 
partially integrated within a few $T$ around $\mu$. The advantage of 
studying this quantity, as compared to the actual interacting DOS 
$\tilde{\rho}_i(\omega)$, is that it can be evaluated without extrapolating the 
Green's function from Matsubara frequencies toward the real energy axis. 

We also note that the weight function $F(\omega)$ in 
$\bar\rho_i(\mu) = -G_{ii}(\beta/2)$ is closely related to the one appearing 
in the conductance of single atoms attached to leads.
In this case the weight function is given by\cite{Meir_Wingreen}
$F_c(\omega)= -{\partial f}/{\partial\omega}$ with halfwidth 
$w_c=2\ln (3+\sqrt{2})T = 3.525 T$. 
Thus, $\bar\rho_i(\mu)$ samples the molecular density of states near $\mu$ 
in a window $\sim 1.5$ times larger than in the case of the conductance.      
 
We point out here that one can, of course, also evaluate the molecular Green's 
function $G_{ij}^{cl}$ close to the real-energy axis. However, 
these spectra consist of many sharp peaks related to the finite number
of levels of the lead clusters. 
Moreover, the level energies $\epsilon_k$ and intra-cluster
coupling terms $v_k$ depend not only on the cluster size, but also on the choice
of the weight function $W_n$ in Eq.\ (\ref{eq10}). Evidently, one would have to
use very large lead clusters so that these discrete spectra evolve into a
meaningful representation of the
continuous spectra of the actual semi-infinite leads. Because of the exponentially
growing Hilbert space the ED approach would then no longer be practical. Thus, the
purpose of introducing the auxiliary fit parameters $\epsilon_k$ and $v_k$ is to
simulate the finite-temperature lead Green's functions and embedding potentials
at Matsubara points. At not too low $T$, these functions converge very well with
cluster size and are remarkably stable against variations in $\epsilon_k$ and $v_k$
and for different choices of $W_n$.\cite{Liebsch:12}    
The continuous spectra of the Green's function elements $G_{ij}(\omega)$ describing
the molecule attached to semi-infinite leads may then be derived via analytic
continuation of $G_{ij}(i\omega_n)$ or, preferably, of the molecular self-energy
$\Sigma_{ij}(i\omega_n)$ to real energies. In the latter case, continuation of the
known one-electron properties of the molecule and of the leads is avoided.
The task of analytically continuing $G_{ij}(i\omega_n)$ or $\Sigma_{ij}(i\omega_n)$
is entirely analogous to the one in quantum Monte Carlo simulations, where the
maximum entropy scheme is often used to generate real energy spectra. 
In DMFT studies the discrete Green's functions $G_{ij}^{cl}(\omega)$ can, however, 
be very useful 
for the identification of a Mott transition since the excitation gap opens at the 
same critical Coulomb interaction as in the continuous spectrum of $G_{ij}(\omega)$.

\section{Results and discussion}
\label{sec_3}

\subsection{Non-interacting molecules}
\label{sec_3A}

\begin{figure}[t]  
\begin{center}
\includegraphics[width=0.4\textwidth]{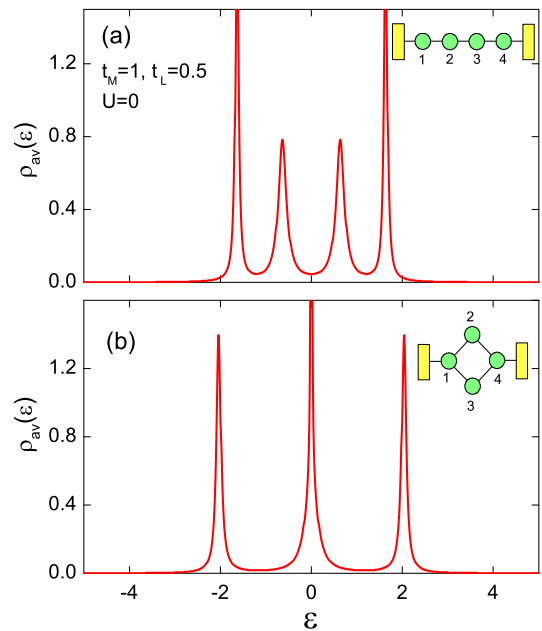}
\end{center}
\caption{\label{Fig_dos}
Non-interacting local DOS averaged over all sites for (a) linear and (b) 
ring molecules ($N=4$) between two semi-infinite leads. The molecular configurations 
are depicted in the insets. $t_M=1$, $t_L=0.5$, and $U=0$.
Imaginary part of energy, $\gamma=0.01$}
\end{figure}

To provide an impression of the electronic structure of the molecule in the 
absence of Coulomb interactions, we show first in Fig.\ \ref{Fig_dos}  
$\rho_{av}(\epsilon)$, the non-interacting local DOS averaged
over all sites, for linear and ring molecules with $N=4$ linked between two
semi-infinite leads, where $t_M=1$  and $t_L=0.5$. In the following, the onsite
energy $\epsilon_a$ is specified as $-U/2$, so that the system becomes electron-hole 
symmetric when it is half-filled. For linear
molecules, the DOS consists of $N$ resonant peaks corresponding to the energy levels 
$e_m$ of the isolated molecule, which are distributed symmetrically with respect to
$\epsilon=0$. The lowest energy state has even parity with respect to the center of
gravity of the molecule, and the parity alternates in the order of ascending energy
levels.  The energy width of the $m$-th resonance, which is determined by its coupling
to the imaginary part of the lead embedding potentials, is given by
\begin{equation}
\Gamma_m = \pi t_L^2 \left[ \rho_0(e_m) |\psi_m(1)|^2+
\rho_{N+1}(e_m) |\psi_m(N)|^2 \right], \label{eq11}
\end{equation}
where $\psi_m(i)$ denotes the amplitude at site $i$ of the electron wave function of
the $m$-th level of the isolated molecule with energy $e_m$. It is understood that
the lowest level corresponds to index $m=1$.
For the ring molecule in Fig.\ \ref{Fig_dos}(b), the DOS exhibits only three peaks,
since the central one at $\epsilon=0$ is doubly degenerate. The wave functions of
the isolated ring molecule in the site basis are,
\begin{eqnarray}
|\psi_1\rangle &=& \frac{1}{2}(|1\rangle+|2\rangle+|3\rangle+|4\rangle),
\nonumber\\
|\psi_2\rangle &=& \frac{1}{\sqrt{2}}(|2\rangle-|3\rangle), \nonumber\\
|\psi_3\rangle &=& \frac{1}{\sqrt{2}}(|1\rangle-|4\rangle), \nonumber\\
|\psi_4\rangle &=& \frac{1}{2}(|1\rangle-|2\rangle-|3\rangle+|4\rangle),
\label{eq12}
\end{eqnarray}
among which $\psi_2$ and $\psi_3$ are degenerate at $\epsilon=0$. Since $\psi_2$
has no amplitude on sites 1 and 4 which are coupled to the leads, $\psi_2$ remains
a truly localized interface state, even when the coupling to the leads is introduced,
so that it makes a $\delta$-function contribution to $\rho_{av}(\epsilon)$.
To avoid this singularity, an artificial imaginary energy $\gamma=0.01$ is used 
in Fig.\ \ref{Fig_dos} for this level. The widths of the other levels correspond
to the physical broadening. It should be noted that, once $U$ is switched on, 
$\psi_2$ and $\psi_3$ are mixed, so that both states become delocalized.

\subsection{Coulomb blockade}
\label{sec_3B}

It should be emphasized that, in contrast to bulk systems, where the ratio $U/t$
(or $U/W$) provides a measure of the strength of 
electron correlations, in the present case an important parameter characterizing
the electronic structure of the interacting molecule in the vicinity of molecular
level $e_m$ is the ratio $U/\Gamma_m$, where $\Gamma_m$ is the level width defined 
in Eq.~(\ref{eq11}).
For $U/\Gamma_m\ll 1$, the molecule is expected to be in the ballistic regime where
correlation effects are dominated by the molecule--lead hybridization.
On the other hand, for $U/\Gamma_m\gg 1$, the molecule is in the Coulomb blockade
regime where the onsite Coulomb repulsion hinders the addition of the second
electron to the molecule when the first one occupies a resonant level.

\begin{figure}[t] 
\begin{center}
\includegraphics[width=0.45\textwidth]{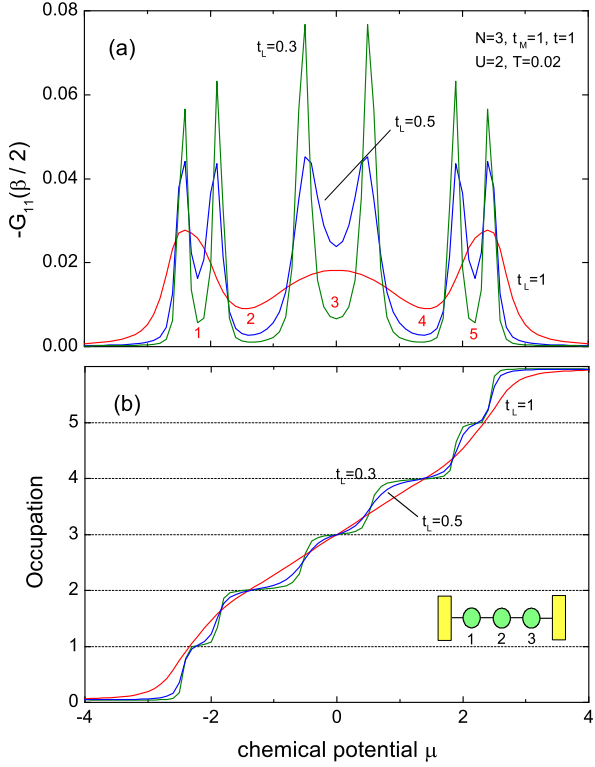}
\end{center}
\caption{\label{Fig_cb1}
(Color online) (a) Partially integrated quasiparticle DOS, 
$\bar\rho_1(\mu)=-G_{11}(\beta/2)$, of a linear molecule with $N=3$ as a function 
of chemical potential $\mu$ for $t_L=0.3$, 0.5 and 1.0 ($U=2$, $t_M=1$, and $T=0.02$).
Small numbers near the minima of these curves indicate chemical potentials giving
integer occupations. 
(b) Electron occupation for the same parameter set as in panel (a).}
\end{figure}

Figure \ref{Fig_cb1}(a) shows $\bar\rho_{1 (3)}(\mu)= -G_{11 (33)}(\beta/2)$
for a chain molecule
($N=3$) as a function of $\mu$ for three values of the contact integral $t_L$.
The other parameters, $t_M=1$, $U=2$, and $T=0.02$, are common to all curves.
The integrated quasiparticle DOS of site 2, $\bar\rho_2(\mu)=-G_{22}(\beta/2)$,
(not shown) is similar to $-G_{11}(\beta/2)$, except that the peak structure 
arising from the second energy level
around $\mu=0$ is absent. Since this molecular orbital has odd
parity with respect to the center of the molecule it has no weight on site 2
irrespective of the magnitude of $U$.
For $t_L=1$, intra-molecular correlation effects are dominated by single-particle
hybridization with the leads. Thus, $\bar\rho_1(\mu)=-G_{11}(\beta/2)$
exhibits three broad peaks as a function of $\mu$, which originate from the energy levels
of the non-interacting molecule. With decreasing $t_L$ (increasing $U/\Gamma_m$,
see Eq.~(\ref{eq11})), the DOS
peaks start exhibiting minima at their centers. For $t_L=0.5$ and 0.3, all three 
peaks split into pairs of peaks separated by a pseudogap induced by Coulomb blockade.
Note that, upon decreasing $t_L$, the double peaks on both sides of the minima 
become sharper, while the energy separation between them depends only weakly on $t_L$.

To analyze this trend, it is useful to expand the Coulomb repulsion in Eq.\ (\ref{eq1})
in terms of the orbital basis of the isolated molecule. Specifically, the 
density--density interaction components are given by
\begin{equation}
\hat{H}_d=\sum_{m,n} U_{mn} \hat{n}_{\psi_m \uparrow} \hat{n}_{\psi_n \downarrow},
\label{eq15}
\end{equation}
where $\hat{n}_{\psi_m \sigma}$ denotes the occupation of the $m$-th orbital with spin
$\sigma$. For the linear molecule with $N=3$, one has
$U_{11}=U_{33}=3U/8$, $U_{22}=U/2$, $U_{12}=U_{23}=U/4$, and $U_{13}=3U/8$.
The effect of other non-diagonal elements not included in Eq.\ (\ref{eq15}) is small
if $U/t$ is not large. In the limit of small $t_L$, the unrestricted self-consistent
field (USCF) approximation may then be used to estimate the mean-field values of the
$m$-th molecular level with spin $\sigma$:
\begin{equation}
\tilde{e}_{m\sigma} = e_m +\sum_n U_{mn}\langle\hat{n}_{\psi_n,-\sigma}\rangle.
\label{eq16}
\end{equation}
Thus, the first molecular level yields peaks at $-\sqrt{2}t-U/2$ and $-\sqrt{2}t-U/8$,
with a gap $3U/8$. The second level has peaks at $-U/4$ and $U/4$, with a gap $U/2$,
while those of the third level are located at $\sqrt{2}t+U/8$ and $\sqrt{2}t+U/2$,
with a gap $3U/8$. The energy positions of the DOS peaks in Fig.\ \ref{Fig_cb1}(a)
are seen to be in fair agreement with these estimates.

Figure \ref{Fig_cb1}(b) shows the total electron occupation of the molecule, summed
over spin and site components, as a function of $\mu$. For $t_L=1$, the occupation
varies smoothly from zero to six. In contrast, for $t_L=0.3$ it exhibits distinct
plateaus at each integer occupation, where those corresponding to odd integers are
caused by the Coulomb blockade effect and their energy positions coincide with those
of the pseudogaps shown in panel (a).

\begin{figure}[t] 
\begin{center}
\includegraphics[width=0.4\textwidth]{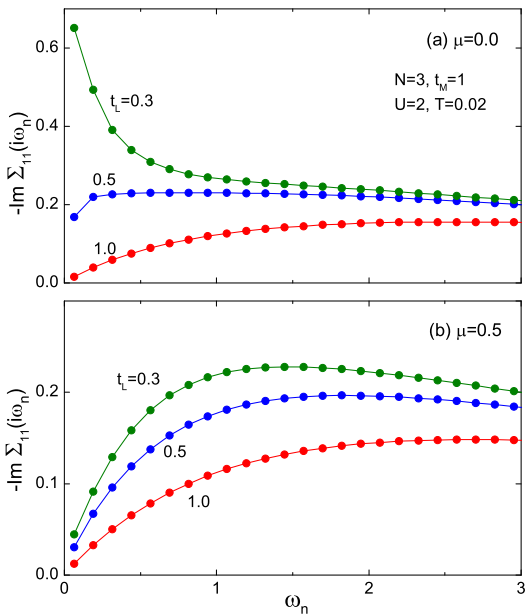}
\end{center}
\caption{\label{Fig_cb2}
(Color online) Imaginary part of correlation-induced self-energy,
$-{\rm Im}\Sigma_{11}(i\omega_n)$, for a linear molecule with $N=3$ at (a) $\mu=0$
(half-filling) and (b) $\mu=0.5$, for three values of molecule-lead coupling $t_L$.
$U=2$, $t_M=1$, and $T=0.02$.}
\end{figure}

To illustrate the effect of Coulomb correlations in the vicinity of the pseudogap,
we plot in Fig.\ \ref{Fig_cb2} the imaginary part of the diagonal element of the 
self-energy, ${\rm Im}\Sigma_{11}(i\omega_n)$, for a chain molecule with $N=3$.
The three curves correspond to $t_L=1$, 0.5, and 0.3. The other parameters are 
the same as in Fig.\ \ref{Fig_cb1}.
Panel (a) shows the self-energy at half-filling ($\mu=0$). For $t_L=1$,
the system is Fermi-liquid-like since ${\rm Im}\Sigma_{11}(i\omega_n)$ tends
to zero linearly as $\omega_n\to 0$. This behavior is in accord with the shape of
the corresponding partially integrated DOS, $\bar\rho_1(\mu)=-G_{11}(\beta/2)$,
shown in Fig.\ \ref{Fig_cb1}(a), which exhibits a quasi-particle peak at $\mu=0$.
As discussed above, with decreasing $t_L$, a Coulomb pseudogap centered at $\mu=0$ 
begins to be formed. As a consequence, at $t_L=0.5$, ${\rm Im}\Sigma_{11}(i\omega_n)$ 
exhibits a finite value at small $\omega_n$, indicating that electrons at
the chemical potential have a finite lifetime inside the molecule at $T=0.02$. 
Upon decreasing $t_L$ further, ${\rm Im}\Sigma_{11}(i\omega_n)$ begins to approach 
the $1/\omega_n$ divergent behavior at small $\omega_n$,
which corresponds to the limit of an isolated molecule.

For comparison, panel (b) illustrates the self-energy at $\mu=0.5$, where the DOS 
exhibits a peak even at small values of $t_L$ (see $-G_{11}(\beta/2)$ in Fig.\ 
\ref{Fig_cb1}(a) for $t_L\le 0.5$). In this case, ${\rm Im}\Sigma_{11}(i\omega_n)$
remains approximately linear in $\omega_n$ even at $t_L=0.3$, indicating that,
outside the pseudogap region, the molecule maintains Fermi-liquid behavior.
Eventually, of course, in the limit $t_L\rightarrow 0$, the metallic behavior
breaks down when the molecule no longer hybridizes with the leads.

\begin{figure}[t] 
\begin{center}
\includegraphics[width=0.4\textwidth]{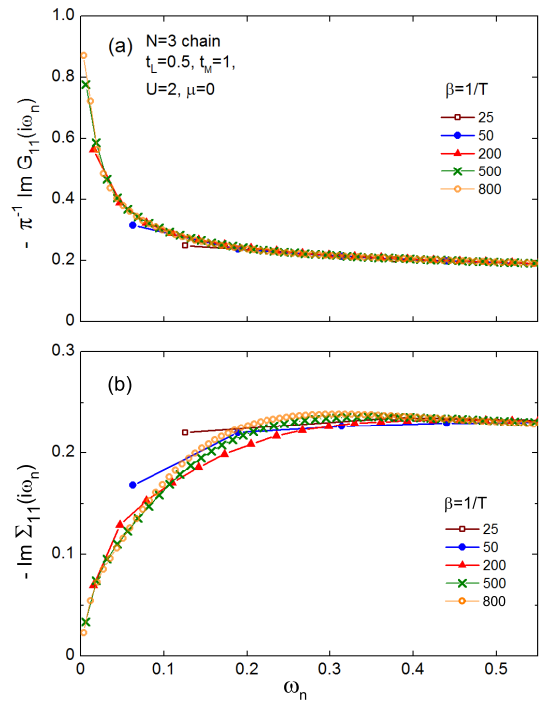}
\end{center}
\caption{\label{Fig_kd1}
(Color online) Imaginary part of (a) interacting Green's function,
${\rm Im}G_{11}(i\omega_n)$, and (b) correlation-induced self-energy,
${\rm Im}\Sigma_{11}(i\omega_n)$, for a linear molecule with $N=3$ for
four inverse temperature values, $\beta=25$, 50, 200, 500, and 800.
$U=2$, $t_L=0.5$, $t_M=1$, and $\mu=0.0$ (half-filling).}
\end{figure}

\subsection{Kondo effect}
\label{sec_3C}

We now discuss the temperature dependence of the electronic structure of the $N=3$ 
chain molecule.  As shown in the Appendix, a cluster consisting of 5 levels 
adequately simulates a semi-infinite lead down to $T\approx 1/800$. Since the 
embedding potential is not affected by the size of the molecule, we use $M=5$ 
lead clusters to investigate the molecular correlation effects in a wide range 
of temperatures.   
Fig.\ \ref{Fig_kd1} shows the imaginary part of the $\{11\}$ element of the 
interacting Green's function and the self-energy as a function of Matsubara 
frequency for $T=1/25, \ldots, 1/800$ and $t_L=0.5$ at half-filling. 
The other parameters are the same as in Fig.\ \ref{Fig_cb2}(a). 
We note here that, in the limit of small $\omega_n$,  
$-\pi^{-1}{\rm Im}G_{11}(i\omega_n)$ coincides with $\tilde{\rho}_1(\omega=0)$, 
the quasiparticle DOS at the chemical potential at site $1$, as indicated in
Eq.\ (\ref{eq14}).
As shown in  panel (a), a sharp quasiparticle peak is formed at $\omega=0$ when
$T$ decreases below about $1/500$. Evidently, this peak may be identified as the 
Kondo resonance caused by the coupling between the localized spin in the 
half-filled second molecular level and the conduction electrons in the leads. 
The behavior of the self-energy is consistent with this trend, as shown in 
panel (b). While $-{\rm Im}\Sigma_{11}(i\omega_n)$ for $T=0.02$ and $0.04$ 
extrapolates to a finite value in the limit of $\omega_n\to0$, at lower $T$ 
it becomes linear in $\omega_n$. Thus, with decreasing $T$ the system undergoes 
a transition from the Coulomb blockade regime to the Kondo regime. 
 
Since the second molecular level is energetically well separated from levels 
1 and 3, its Kondo temperature may be estimated as follows.
The wave function of this level in the site basis is given by
\begin{equation}
|\psi_2\rangle=\frac{1}{\sqrt{2}}\left(|1\rangle-|3\rangle\right) \nonumber
\end{equation}
and the hybridization strength defined in Eq.\ (\ref{eq11}) is
\begin{equation}
\Gamma_2=\pi t_L^2 [\rho_0(0)+\rho_4(0)]/2=\pi t_L^2 \rho_0(0)=0.131.
\nonumber
\end{equation}
Moreover, as pointed out above, the effective Coulomb interaction for this
level is $u=U_{22}=U/2=1$. Thus, within the context of the half-filled Anderson 
model, we have $u/\Delta \gg 1$, where $\Delta$ corresponds to $\Gamma_2=0.131$. 
The Kondo temperature is then approximately given by the expression
\cite{Hewson:93}
\begin{equation} 
  T_K = (u {\Delta}/2)^{1/2} e^{-{\pi u}/{8\Delta} + {\pi\Delta}/{2u}} = 0.0157 .
\label{eq17}
\end{equation}
This estimate is fully consistent with the results in Fig.~\ref{Fig_kd1}, where 
both the Green's function and correlation-induced self-energy exhibit no 
noticeable changes as a function of temperature for $T\alt T_K$. Moreover, from the 
initial slope of Im\,$\Sigma_{11}$ we obtain a quasiparticle weight $Z\approx 1/8$, 
yielding\cite{Hewson:05} 
$T_K = \pi Z \Delta/4 \approx 0.013$, in reasonable agreement with the 
estimate quoted above.      

\begin{figure}[t] 
\begin{center}
\includegraphics[width=0.4\textwidth]{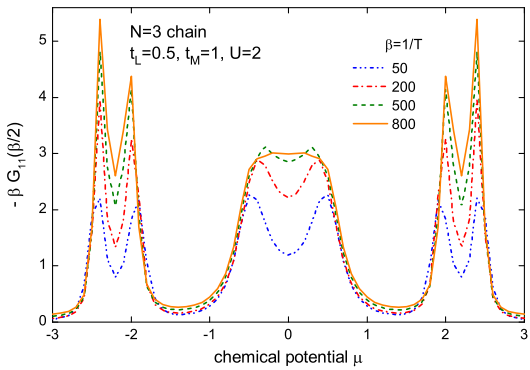}
\end{center}
\caption{\label{Fig_kd2}
(Color online) 
Partially integrated quasiparticle DOS, $\bar\rho_1(\mu)=-G_{11}(\beta/2)$ 
(multiplied by $\beta$), of a linear molecule with $N=3$ as a function of 
chemical potential $\mu$ for $T=1/50$, 1/200, 1/500, and 1/800.
$t_L=0.5$, $t_M=1$, and $U=2$.}
\end{figure}

Figure \ref{Fig_kd2} shows the temperature dependence of the partially
integrated quasiparticle DOS at site 1, $\bar\rho_1(\mu)=-G_{11}(\beta/2)$, 
for the same molecule as in Fig.\ \ref{Fig_kd1}. To compare different
temperatures, the curves are multiplied by $\beta$ since the weight function 
$F(\omega)$ in Eq.\ (\ref{eq15n}) has integrated weight $\pi T$.    
As discussed above, the Kondo resonance appears at $\mu$ for $T\alt T_K$,
where $T_K$ depends on $\mu$. Apparently, $T_K$ associated with the second 
molecular level is larger than or comparable to the lowest value $T=1/800$ in
Fig.\ \ref{Fig_kd2}. Since the Kondo resonance appears at an energy close to 
$\mu=0$, it makes a large contribution to $\bar{\rho}_1(\mu)$.
As a consequence, the minimum (Coulomb pseudogap) between the two peaks for 
the second molecular level for $T=0.02$ becomes shallower with decreasing $T$, 
and is eventually replaced by a broad single peak at $T=1/800$.

Interestingly, in contrast to this behavior of the second molecular level, the 
Coulomb pseudogaps for the first and third levels remain visible for the whole $T$ 
range in Fig.\ \ref{Fig_kd2}. The difference arises from the lower $T_K$ values 
for these levels. The wave function of the first level of the isolated molecule is
\begin{equation}
|\psi_1\rangle=\frac{1}{2}\left(|1\rangle+\sqrt{2}|2\rangle+|3\rangle\right),
\nonumber
\end{equation}
and the hybridization strength is
\begin{equation}
\Gamma_1=\pi t_L^2 [\rho_0(\tilde{e}_1)+\rho_4(\tilde{e}_1)]/4
=\pi t_L^2 \rho_0(\tilde{e}_1)/2=0.046,
\nonumber
\end{equation}
where $\tilde{e}_1$ denotes the centroid of the doublets, $-\sqrt{2}t-5U/16=-2.04$.
By inserting $\Delta=\Gamma_1$ and $u=U_{11}=3U/8=0.75$ in Eq.\ (\ref{eq17}) 
for the case of half-filling, we obtain $T_K=2.4\times 10^{-4}$, which is much lower
than the temperature range in Fig.\ \ref{Fig_kd2}. 

\begin{figure}[t] 
\begin{center}
\includegraphics[width=0.4\textwidth]{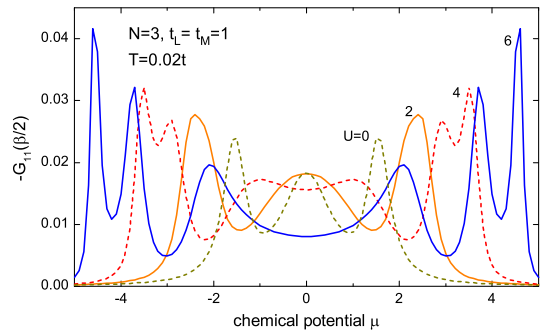}
\end{center}
\caption{\label{Fig_U}
(Color online) Partially integrated quasiparticle DOS, 
$\bar\rho_1(\mu)=-G_{11}(\beta/2)$, of a linear molecule with $N=3$ as a function 
of chemical potential $\mu$ for $U=0$, 2, 4, and 6. $t_L=1$, $t_M=1$, and $T=0.02$.}
\end{figure}

\subsection{Large $U$ region}
\label{sec_3D}

In Fig.\ \ref{Fig_cb1}, we fixed the onsite Coulomb repulsion $U$ and varied the
hopping integral between leads and molecule. Alternatively, it is of interest
to inquire how the molecular electronic structure changes when $U$ is increased 
for fixed hopping. Here we consider the case of strong coupling where $t_L=t_R=1$.
Figure \ref{Fig_U} shows the partially integrated quasiparticle DOS at site 1,
$\bar\rho_1(\mu)=-G_{11}(\beta/2)$, for a chain molecule
with $N=3$ as a function of $\mu$ for four values of $U$. The other parameters
are the same as in Fig.\ \ref{Fig_cb1}. As $T/t_L\ll 1$, for $U=0$, $-G_{11}(\beta/2)$
is nearly identical to the non-interacting local DOS at site 1, $\rho_1(\mu)$,
except for a constant factor. The curve for $U=2$ coincides with the one shown in
Fig.\ \ref{Fig_cb1}(a). Compared with the bare non-interacting DOS, the three peaks
are considerably broadened as a result of the intra- and inter-molecular orbital
Coulomb terms appearing in Eq.\ (\ref{eq15}). Moreover, the outer peaks are
shifted to higher energies relative to the corresponding peaks at $U=0$.
When the Coulomb energy is increased to $U=4$, all three DOS peaks begin to exhibit
a minimum at their center. Finally, they evolve into double peak structures at $U=6$.

Interestingly, the energy separations between the double peaks in Fig.\ \ref{Fig_U}
differ from those in the USCF approximation discussed above ($3U/8=2.25$
for the first and third orbitals and $U/2=3$ for the second orbital). This
indicates that the off-diagonal Coulomb matrix elements ignored in Eq.\ (\ref{eq15})
become progressively more important with increasing $U/t$. Consequently, for $U=6$,
the three orbitals are significantly mixed by these off-diagonal terms. The spectrum
may then more correctly be interpreted in terms of upper and lower Hubbard bands,
each consisting of three peaks and split by the Mott-like gap at the center.
Nevertheless, due to the proximity effect the DOS remains finite even at low
energies because the molecule is strongly coupled to the two metallic leads.
 
\begin{figure}[t] 
\begin{center}
\includegraphics[width=0.4\textwidth]{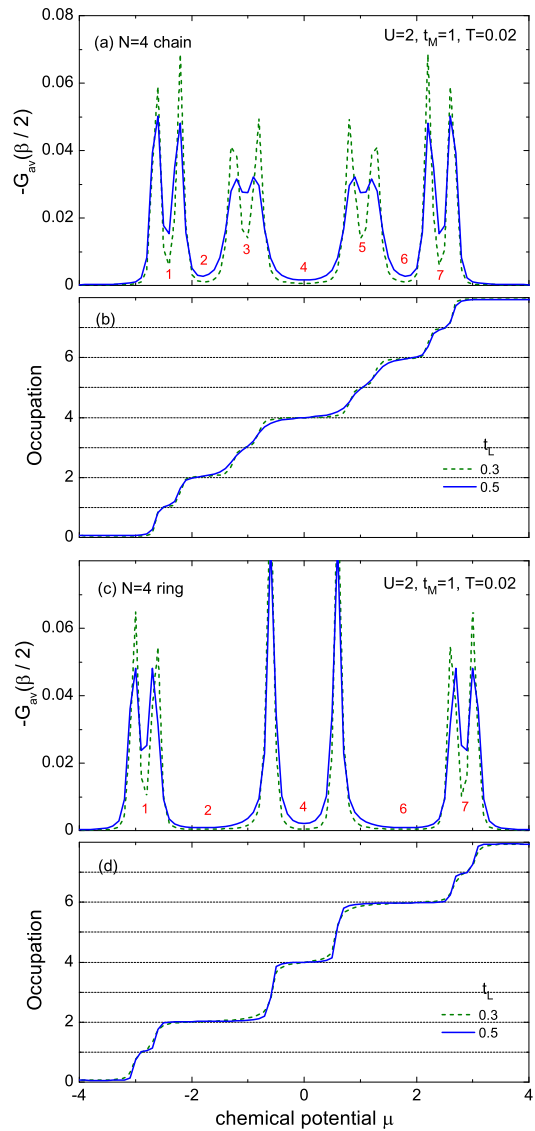}
\end{center}
\caption{\label{Fig_N4}
(Color online) Partially integrated quasiparticle DOS averaged over all the sites,
$-G_{av}(\beta/2)$, for a (a) linear and (c) ring molecule with $N=4$ for
$U=2$, $t_M=1$, and $T=0.02$. Solid (blue) and dashed (green) lines correspond 
to $t_L=0.5$ and
$t_L=0.3$, respectively. The corresponding local densities of states for the
non-interacting molecules ($U=0$) are shown in Fig.\ \ref{Fig_dos}.
Small numbers near the minima of these curves indicate chemical potentials giving
integer occupations. 
Panels (b) and (d) provide the corresponding occupancies as a function of $\mu$. 
}
\end{figure}

\subsection{Chain vs. ring molecules}
\label{sec_3E}

So far we have presented results for a chain molecule with $N=3$. The results for
other linear molecules that we have studied ($N=2$ to 5) are qualitatively similar
and can be summarized as follows:
(i) For a weakly correlated molecule, the quasiparticle DOS at the chemical potential
$\mu$ exhibits $N$ peaks corresponding to the $N$ energy levels $e_m$ of the molecule,
whose width is determined by the hopping integral between the lead and molecule, $t_L$. 
(ii) With increasing $U/\Gamma_m$, these peaks start exhibiting minima at their centers.
(iii) When $U/\Gamma_m$ is increased further, each quasiparticle DOS peak becomes a
double peak structure, so that $\bar\rho_i(\mu)=-G_{ii}(\beta/2)$ consists of $2N$
quasiparticle peaks as a function of $\mu$, instead of the $N$ peaks in the
non-interacting limit.
In the range of the correlation-induced pseudogaps, the electron self-energy exhibits
a finite scattering rate. Furthermore, the electron occupation of the molecule
as a function of $\mu$ exhibits plateaus at odd integers, whose energy positions
correspond to the location of the Coulomb pseudogaps. (iv) When temperatures is
lowered to reach $T_K$, which depends strongly on the Hamiltonian parameters and the
molecular levels, Kondo resonances are formed in the Coulomb pseudogaps as a result
of the strong coupling of the localized spin and conduction electrons in the leads. 
 
As an example, we plot in Fig.\ \ref{Fig_N4}(a) the quantity $-G_{av}(\beta/2)$, 
which is defined as the average of $\bar\rho_i(\mu)=-G_{ii}(\beta/2)$ over the $N$ 
molecular sites, for a linear molecule with $N=4$ for two values of the molecule-lead 
coupling parameter, $t_L=0.3$ and 0.5. The other parameters, $U=2$, $t_M=1$, and 
$T=0.05$, are the same as in Fig.\ \ref{Fig_cb1}.
The comparison of Figs.\ \ref{Fig_dos} and \ref{Fig_N4} illustrates how the
quasiparticle DOS peaks at the chemical potential $\mu$ evolve when the onsite
Coulomb energy is increased from zero to a finite value.
As mentioned above, there is a one-to-one correspondence between the DOS peaks
in Fig.\ \ref{Fig_dos} and the double peak structures in Fig.\ \ref{Fig_N4}.
For the present molecule, the intra- and inter-molecular-orbital Coulomb energies
are calculated as: $U_{mm}=3U/10$ ($m=1$ to 4), $U_{12}=U_{13}=U_{24}=U_{34}=U/5$,
and $U_{14}=U_{23}=3U/10$. Thus, at positive $\mu$, the DOS peaks of the third
molecular orbital within the USCF approximation and in the small $t_L$
limit are located at $(\sqrt{5}-1)t/2$ and $(\sqrt{5}-1)t/2+3U/10$, with a gap
$3U/10$, while those of the fourth level appear at $(\sqrt{5}+1)t/2+U/5$ and
$(\sqrt{5}+1)t/2+U/2$, also with a gap $3U/10$. The peaks at negative $\mu$ are
located symmetrically with respect to $\mu=0$. The Coulomb pseudogaps in
Fig.\ \ref{Fig_N4}(a) are in fair agreement with these mean-field values.

Finally, we discuss the ring molecule with $N=4$ since it behaves quite differently 
from the corresponding linear molecule at low temperatures when it is half-filled.
In Fig.\ \ref{Fig_N4}(b) we plot $-G_{av}(\beta/2)$ of this molecule for the
same parameter set as for the linear molecule in Fig.\ \ref{Fig_N4}(a).
With the molecular orbitals defined in Eq.\ (\ref{eq12}), the intra-orbital
Coulomb energies are $U_{11}=U_{44}=U/4$, $U_{22}=U_{33}=U/2$, while the
inter-orbital ones are $U_{mn}=U/4$ ($m\ne n$), except for $U_{23}=0$.
Hence, the DOS peaks of the first (fourth) level appear at  $-2t-U/2$ and $-2t-U/4$
($2t+U/4$ and $2t+U/2$), with a gap $U/4$, while the degenerate peaks of the second
and third levels are located at $-U/4$ and $U/4$, with a twice larger gap $U/2$.
The peak energies in Fig.\ \ref{Fig_N4}(b) are in agreement with these USCF
estimates. The ring molecule is half-filled when $\mu$ is located inside the
pseudogap between the doublets at $\mu\sim\pm U/4$,
with the first level essentially fully occupied and with the second and third ones
singly occupied. 

\begin{figure}[t] 
\begin{center}
\includegraphics[width=0.4\textwidth]{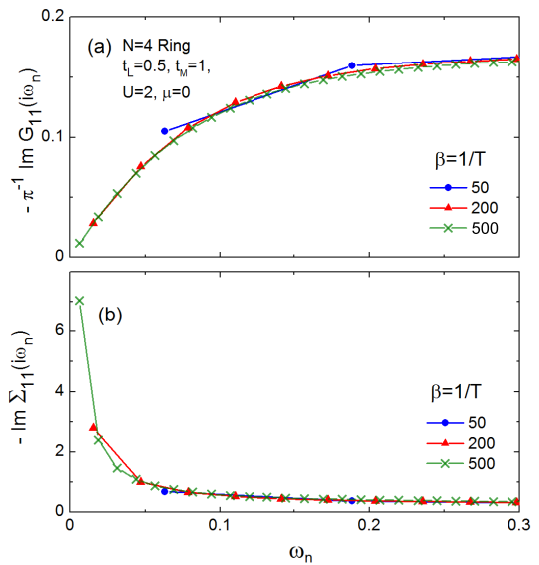}
\end{center}
\caption{\label{Fig_kd3}
(Color online) Imaginary part of (a) interacting Green's function,
${\rm Im}G_{11}(i\omega_n)$, and (b) correlation-induced self-energy,
${\rm Im}\Sigma_{11}(i\omega_n)$, for a ring molecule with $N=4$ for
inverse temperature values, $\beta=50$, 200, and 500.
$U=2$, $t_L=0.5$, $t_M=1$, and $\mu=0.0$ (half-filling).}
\end{figure}

The question arises as to whether the electrons in the second and third levels
form a singlet or triplet state in the many-body ground state
\cite{hofstetter,kogan,roch} which may arise as a consequence of off-diagonal
Coulomb matrix elements neglected in Eq.\ (\ref{eq15}).
For the isolated molecule with the same $U$, we found that the singlet state has 
a lower energy. Thus, because of the absence of a localized-spin degree of freedom, 
the $N=4$ ring molecule at half-filling exhibits no Kondo proximity effect.
To confirm this, we plot in Fig.\ \ref{Fig_kd3} the $\{11\}$ element of the 
interacting Green's function and self-energy at half-filling. 
In striking contrast to Fig.\ \ref{Fig_kd1} for the linear molecule with $N=3$, 
the correlation-induced self-energy is seen to preserve non-Fermi-liquid behavior 
at low temperatures, and $\tilde{\rho}(\omega=0)$, i.e., the low frequency limit 
of $-\pi^{-1}{\rm Im}G_{11}(i\omega_n)$ in panel (a) does not exhibit a Kondo
resonance.  This explains why
$-G_{av}(\beta/2)$ in Fig.\ \ref{Fig_N4}(b), when $\mu$ is located inside the
pseudogap ($|\mu|\alt U/4$), is much smaller than the corresponding one for
the linear molecule with $N=3$ shown in Fig.\ \ref{Fig_cb1}(a), despite the
fact that the pseudogap is nearly the same ($\sim U/2$) for both molecules.

\section{Summary}
\label{sec_4}

A new method for the evaluation of the electronic properties of strongly correlated 
molecules coupled to semi-infinite metallic leads is proposed. By simulating the surface 
Green's functions of the leads in terms of small clusters, the many-body interactions
of the combined system in the zero bias-voltage limit are obtained via exact 
diagonalization. The auxiliary energies and hopping terms of the lead clusters 
are derived by fitting the lead surface Green's functions at imaginary Matsubara 
frequencies. These fits are found to be sufficiently accurate to describe the Kondo 
physics, except in the limit of extremely low temperatures. 
For moderate onsite Coulomb energies within the molecule, the density of states
peaks of the non-interacting molecule are shown to split into doublets separated
by correlation-induced pseudogaps. The molecular self-energy then exhibits a finite 
scattering rate, as expected in the regime of Coulomb blockade. 
Outside the pseudogap regions, the self-energy retains
ordinary Fermi-liquid behavior, characteristic of ballistic transport across the 
molecule. The one-electron hybridization between molecule and leads is shown to be 
a key parameter that governs the transition between the ballistic and Coulomb blockade 
regimes. If the chemical potential is located inside a pseudogap, the molecular 
levels are integer occupied, so that a Kondo resonance appears upon lowering the 
temperature in the case of odd integer oocupancies.  
The present results suggest that the approach discussed in this work for molecules
or quantum dots connected to metallic leads can describe, as a function of 
Hamiltonian parameters, electron filling, and temperature, the full range of 
phenomena from Coulomb blockade to Kondo physics. In future applications it would 
be interesting to apply this scheme to multiorbital dots and a variety of other
models of interest for nanoscale devices.

\begin{acknowledgments}
We like to thank Theo Costi for comments on the manuscript and valuable discussions.
H.I. thanks the Forschungszentrum J\"ulich for support during his stay in Germany.
Part of the computations were carried out at the  J\"ulich Supercomputing Center.
\end{acknowledgments}

\appendix*
\section{Single adatom on a semi-infinite lead}

\begin{figure}[t] 
\begin{center}
\includegraphics[width=0.4\textwidth]{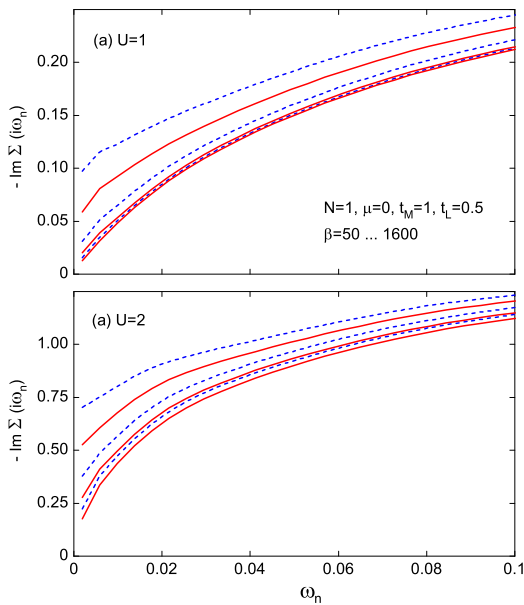}
\end{center}
\caption{\label{FigA1}
(Color online) Imaginary part of self-energy of single atom on semi-infinite 
lead along Matsubara axis at half-filling for temperatures corresponding to 
$\beta=1/T =50,100,200, \dots, 1600$ (from above). 
(a) $U=1$ and (b) $U=2$. $t_M=1$, $t_L=0.5$.
All results are obtained for $M=11$, with a fixed Matsubara grid corresponding 
to $\beta_M=1600$.  
}
\end{figure}

\begin{figure} 
\begin{center}
\includegraphics[width=0.4\textwidth,angle=0]{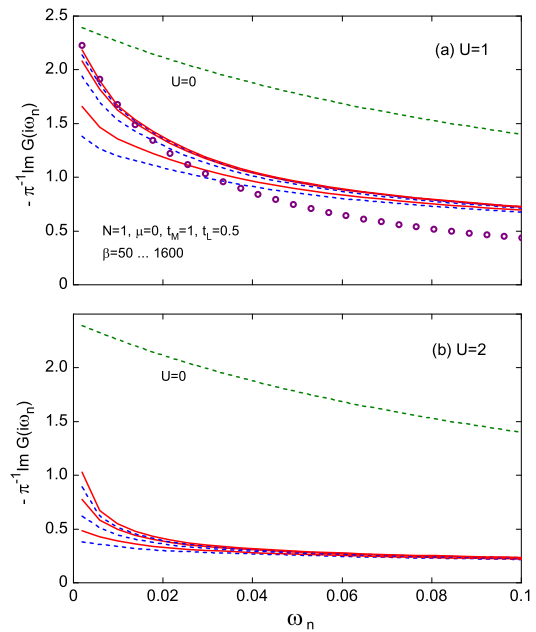}
\end{center}
\caption{\label{FigA2} 
(Color online) Imaginary part of Green's function of single atom on 
semi-infinite lead. The parameters are the same as in Fig. \ref{FigA1}.
The uncorrelated Green's function ($U=0$) is indicated by the green dashed 
curve. The symbols in panel (a) correspond to the Green's function 
$1/(i\omega_n /Z + i\Delta)$ for a single peak at $\mu=0$ with Kondo
temperature $T_K=\pi Z\Delta/4$ (see text).}  
\end{figure}

In Section \ref{sec_2}, we have discussed the main new feature of the present scheme,
namely, the simulation of the semi-infinite leads in terms of a finite set of 
levels. Essentially, the true embedding potentials which have continuous spectra
at real energies are replaced by those for finite clusters comprising a discrete
set of poles. The criterion for this substitution is that along the Matsubara 
axis both versions of the embedding potentials should agree well for a given
cluster size. Evidently, this fitting is accurate only at not too low temperature
when the lowest Matsubara point is not too close to the real energy axis.        
Thus, for each cluster size, there should be a low temperature limit down to 
which the discrete set of levels accurately mimics the electronic properties 
of the actual semi-infinite lead. This limit may be determined by performing 
calculations for clusters with different sizes and by checking the consistency
of the corresponding results.  
 
\begin{figure}[t] 
\begin{center}
\includegraphics[width=0.4\textwidth,angle=0]{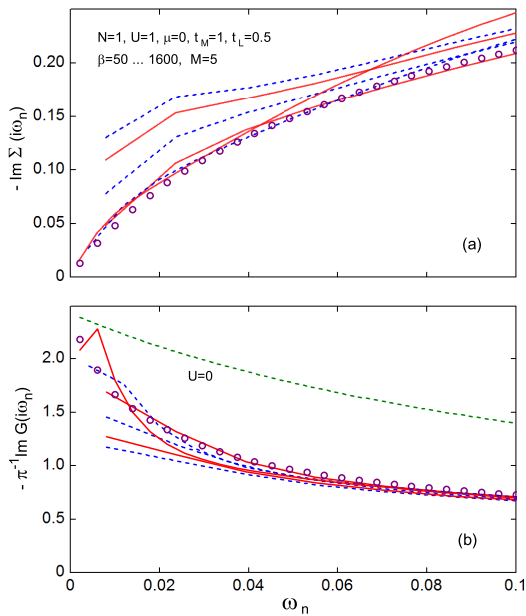}
\end{center}
\caption{\label{FigA3} (Color online) 
Upper (lower) panel: as in upper panel of Fig. \ref{FigA1} (\ref{FigA2}) except 
for $M=5$. The Matsubara grid corresponds to $\beta_M=1600$ for $\beta\ge 400$ 
and to $\beta_M=400$ for $\beta< 400$. The results for $M=11$ at $\beta=1600$ 
are indicated by the circles. 
}\end{figure}

For this purpose we consider here the special case of a single atom ($N=1$) 
adsorbed on a single semi-infinite lead which is equivalent to the single-impurity
Anderson model. The electronic structure of the correlated atom can therefore be
directly compared with predictions within NRG.\cite{Oguri:05,
Numata:09,Tanaka:10,Mitchell:11,Vernek:11,Misiony:12,Costi:10,Hewson:93}   
If the lead is replaced by a cluster containing $M$ levels, the calculation of 
the Green's function of the combined system involves the eigenstates of a 
$(M+1)$-level cluster. To be specific, we choose hopping parameters $t_M=1$ and 
$t_L=0.5$. The Coulomb interaction is assumed to have the values $U=1$ and $U=2$.
For the simple-cubic structure of the lead, the local density of states at $\mu=0$
in the surface layer is $\rho_s(0)=0.52/\pi$, so that the effective hybridization 
between atom and lead is $\Delta = \pi t_L^2\rho_s(0)=0.13$. In Kondo physics, it 
is customary to introduce the parameter $x=U/(\pi\Delta)$ to characterize the
importance of Coulomb repulsion versus single-electron hopping. Thus, for $U=1$
and $U=2$, this parameter has the values $x=2.43$ and $x=4.86$, respectively.     
According to Eq.\ (\ref{eq17}), the Kondo temperature then has the values:  
$T_K=0.0157$ for $U=1$ and $T_K=0.001$ for $U=2$.

As discussed in Ref.~\onlinecite{Liebsch:12}, finite-temperature exact diagonalization
can now be carried out for clusters involving up to about $n_s=15$ levels. Here we 
consider lead clusters up to $M=11$, i.e., $n_s=12$. As shown below, these sizes
are sufficient for temperatures down to about $T=1/1600$, i.e., well within the 
Kondo regime for $U=1$ and above about $0.6\,T_K$ for $U=2$. Increasing $M$ to 14 
would permit the study of even lower temperatures.  

Also, we point out here that, while the fitting of the lead Green's function in 
Eq.\ (\ref{eq10}) is usually done by using the Matsubara points corresponding to 
the physical temperature $T$ in Sec.\ \ref{sec_3}, it is possible to introduce a 
fictitious Matsubara grid 
independently of $T$, which is used only for the purpose of fitting the lead Green's
function, i.e., for finding the parameter set $\{\epsilon_k, v_k\}$. We denote this
fictitious Matsubara temperature by $T_M$. Because of the large cluster size ($M=11$),
we choose $T_M=1/1600$, which should therefore provide excellent fits of $g_{00}$
in the entire range of physical temperatures considered, $T=1/1600,\ldots,1/50$.  
The small value of $T_M$ implies that accurate low-energy behavior of the cluster 
Green's function $G_{ij}^{cl}$ is available. We caution, however, that, in using 
this technique, it is important to check that the cluster Green's function agrees 
with that of the lead not only on the fitting points in Eq.\ (\ref{eq10}), but 
also on the real Matsubara points corresponding to the physical temperature.

Figure \ref{FigA1} shows the low-energy behavior of the self-energy of the adatom 
for various temperatures. For $U=1$ (upper panel) and $T$ approximately less than 
$0.002$ ($\beta > 500$), Im\,$\Sigma(i\omega_n)$ is linear $\sim i \omega_n$, 
with a slope of about $6.4$, yielding a quasi-particle weight $Z\approx 0.135$.
At larger temperature, the self-energy develops a finite onset, indicating a growing
correlation induced low-energy scattering rate. The finite lifetime at $\mu=0$ 
then implies that the pinning condition of the interacting density of states is 
increasingly violated. From the initial slope of the self-energy we can estimate 
the Kondo temperature by using the expression\cite{Hewson:05} 
$T_K = \pi Z \Delta /4$.  Thus, $T_K\approx 0.014$  for $U=1$, which agrees well 
with the estimate $T_K=0.0157$ derived from Eq.~(\ref{eq17}). According to the lower 
panel, for $U=2$ the linear region of the self-energy is confined to much lower 
temperatures, with a quasi-particle weight of about $Z \alt 0.01$ and $T_K \alt 0.001$. 
This result is also in agreement with the estimate obtained from Eq.~(\ref{eq17}).

Figure \ref{FigA2} shows the quasi-particle Green's function for the correlated 
adatom in the same temperature range as in Fig.~\ref{FigA1}. For $U=1$ and 
$T<0.002$, Im\,$G(i\omega_n)$ is seen to approach the same low-energy limit 
as the uncorrelated Green's function, as expected from the pinning condition. 
With increasing $T$, deviations from this condition become progressively larger, 
in correspondence with the behavior of the self-energy shown in Fig.~\ref{FigA1}. 
For $U=2$, extrapolation of Im\,$G(i\omega\rightarrow 0)$ to the pinning condition 
might be feasible only at the lowest temperature, in agreement with the estimate 
$T_K\approx 0.001$ given above. The behavior of $G(i\omega)$ in the very-low-$T$ 
region could be explored with greater accuracy by enlarging the lead cluster 
beyond $M=11$.          
 
To illustrate the accuracy of the adatom self-energy and Green's function for 
smaller lead clusters, we show in Fig.~\ref{FigA3} the results for $U=1$ and $M=5$.
For $\beta=1600$, a comparison with the results for $M=11$ is also provided.  
Although the Green's function now is less accurate at low energies because
of the reduced number of cluster levels, the self-energy still exhibits the 
correct qualitative trend: Below about $T \le 1/400$, Im\,$\Sigma(i\omega_n)$
is linear in $\omega_n$, whereas at larger $T$ the low-energy scattering rate  
increases significantly. At low $T$, the initial slope of Im\,$\Sigma(i\omega_n)$
is nearly the same as for $M=11$, yielding similar quasiparticle weight and 
Kondo temperature. Thus, in spite of the larger quantitative uncertainties
in the case of the smaller lead cluster, the overall evolution, namely, from 
Fermi liquid behavior at low $T$ to increasing low-energy scattering rates 
beyond about $T=1/400$, is consistent with the more precise results for $M=11$.

\end{document}